\documentclass[fleqn,usenatbib]{mnras}
\usepackage{natbib}
\usepackage{deluxetable}

\usepackage{newtxtext,newtxmath}

\usepackage[T1]{fontenc}
\usepackage{ae,aecompl}

\usepackage{graphicx}	
\usepackage{amsmath}	
\usepackage{amssymb}	

\newcommand{\dlndt}[1]{\dfrac{\dot{#1}}{#1}}
\newcommand{\fqg}{f\left(q,\gamma\right)}
\newcommand{\id}{{\rm d}}

\title[GW-Moderated Accretion]{Gravitational-wave-moderated Accretion:\\ The Case of ES Ceti}

\author[M. S. B. Coleman et al.]{
Matthew S. B. Coleman,$^{1}$\thanks{E-mail: mcoleman@ias.edu (MSBC)}
Tejaswi Venumadhav,$^{1}$
and Barak Zackay$^{1}$
\\
$^{1}$Institute for Advanced Study, Einstein Drive, Princeton, NJ 08540, USA
}

\date{Accepted XXX. Received YYY; in original form ZZZ}

\pubyear{2019}

\begin{document}
\label{firstpage}
\pagerange{\pageref{firstpage}--\pageref{lastpage}}
\maketitle

\begin{abstract}
We show that recent observations of the compact binary, AM~CVn type system, ES Ceti are fully consistent with theoretical predictions of stable mass transfer moderated by angular momentum loss due to gravitational-wave radiation. 
One of the main predictions of this model (for degenerate donors) is a widening of the binary. 
The mass transfer rate inferred from the observed rate of change in the orbital frequency is consistent with that inferred from the observed flux using the recent Gaia DR2 parallax. 
\end{abstract}




\section{Introduction}


Ultracompact binaries with orbital periods $\lesssim 20$ min are expected to be strong emitters of gravitational waves, and are one of the prime targets for the planned space-based Laser Interferometer Space Antenna (LISA) mission. Many of these binaries will be individually resolvable by LISA \citep{NEL04,KUP18}; of the known ultracompact binaries $\sim16$ will be detectable with LISA (so-called verification binaries), most of which are semi-detached white dwarf binaries classified as AM Canum Venaticorum type stars \citep[AM~CVns;][]{KUP18}.
These very short period ($\lesssim 65$ minutes) binary systems consist of a white dwarf accreting from a hydrogen-deficient companion star (often another, lower mass, white dwarf) through Roche lobe overflow, and are sometimes referred to as
helium cataclysmic variables (see, e.g.,
\citealt{SOL10} for a review).

In addition to their gravitational radiation, AM~CVns are interesting astrophysical objects for a variety of reasons. The discs in these systems feature all the rich
phenomenology of hydrogen-dominated white dwarf
accretion discs, making them useful laboratories for studying the effects of composition on accretion physics \citep[see e.g.][]{COL18}. AM~CVns may also be sources of helium novae, type Ia supernovae, and
.Ia supernovae \citep{BIL07}.

As noted by \citet{P67}\footnote{\citet{P67} discussed AM~CVns when only the prototype was known (AM~CVn; then called HZ~29).}, observations of the orbital evolution of AM~CVns can be used to confirm the effects of gravitational radiation. For conservatively mass transferring systems (where the total mass is conserved) with a sufficiently degenerate donor, emission of gravitational radiation causes the orbit to widen rather than shrink. We refer to this effect as gravitational-wave-moderated accretion.
In order to test this model, orbital evolution (i.e. a period derivative) must be detected in a system. To date, only one AM~CVn with an accretion disc has a detected positive period derivative: ES Ceti \citep{fdot_ES_Ceti}. 

Here, we focus on the orbital evolution of AM~CVns as a means to infer their emission of gravitational waves. 
In section 2 we derive and discuss our model for orbital evolution of gravitational-wave-moderated mass transferring binaries, some of which is a review of \citet{P67} and more recently \citet{2004MNRAS.350..113M}.
In section 3 we apply this model to the system ES Ceti.
We discuss the implications of our
results in section 4, 
and summarise our conclusions in section 5.

\section{Model}
\label{sec:model} 



We consider a (semi-detached) mass transferring binary system with an accretor mass of $M_{\rm a}$ and a donor mass of $M_{\rm d}$, with a mass ratio of $q\equiv M_{\rm d}/M_{\rm a}$. We adopt the following assumptions:
\begin{enumerate}
    \item The mass transfer is conservative, i.e. $\dot{M}\equiv\dot{M}_{\rm a}=-\dot{M}_{\rm d}$.
    \item The donor fills its Roche lobe, i.e., its radius $R_{\rm d}$ equals the Roche radius $R_{\rm L}$.
    \item Gravitational waves are the only angular momentum sink in the system.
\end{enumerate}
We work with the following approximation for the Roche radius $R_{\rm L}$:
\begin{align}
\label{eqn:roche}
\dfrac{R_{\rm L}}{a}=\dfrac{2}{3^{4/3}}\left(\dfrac{q}{1+q}\right)^{1/3}=\eta \tilde{q}^{1/3},
\end{align}
which is accurate to $2\%$ for $0<q<0.3$ \citep{K59,P67}. 
In Eqn.~\eqref{eqn:roche}, we defined $\eta \equiv 2/3^{4/3} \approx 0.46$ and $\tilde{q}\equiv q/(1+q)$ to simplify further expressions.

Using Newton's form of Kepler's third law, the period $P$ of the binary orbit is
\begin{align}
\label{eqn:kep}
P^2=a^3\dfrac{4\pi^2}{G\left(M_{\rm a}+M_{\rm d}\right)} = (a^3 \tilde{q}) \dfrac{4\pi^2}{GM_{\rm d}}.
\end{align}
We plug in Eqn.~\eqref{eqn:roche} for the term in brackets, and use the assumption that the donor exactly fills its Roche lobe ($R_{\rm L}=R_{\rm d}$), to write the period in terms of the mass and radius of the donor
\begin{align}
  P^2 & = \dfrac{4\pi^2}{\eta^3} \dfrac{R_{\rm d}^3}{GM_{\rm d}}. \label{eq:period}
\end{align}
This tells us that the orbital period is closely related to the dynamical time of the donor star.
At first sight, it might appear surprising that the dependence on $q$ falls out; this is due to our choice of the approximation in Eqn.~\ref{eqn:roche} over that of \citet{EGG83}; the latter would lead to a weak $q$ dependence in Eqn.~\eqref{eq:period}.

To make further progress, we take the mass-radius relation of the donor to be a power law
\begin{equation}
\label{eqn:MR}
R_{\rm d}=R_0\left(\dfrac{M_{\rm d}}{M_0}\right)^\gamma,
\end{equation}
where $R_0$, $M_0$, and $\gamma$ are arbitrary constants, which we additionally assume to vary only negligibly with time, if at all. If we define the characteristic timescale 
\begin{equation}\label{eqn:P0}
P_0\equiv\sqrt{\dfrac{4\pi^2R_0^3}{\eta^3 GM_0}} = 523 \, {\rm s} \times \left( \dfrac{R_0}{3\times10^{-2} R_{\sun}} \right)^{3/2} \left( \dfrac{M_0}{0.1 \, M_\odot} \right)^{-1/2},
\end{equation}
we can express the period $P$ as 
\begin{align}
P &= P_0 \left(\dfrac{M_{\rm d}}{M_0}\right)^{(3\gamma-1)/2}. \label{eqn:periodpower}
\end{align}
Turning Eqn.~\eqref{eqn:periodpower} around, we can infer the donor's mass from the orbital period of the binary:
\begin{align}
M_{\rm d} &= M_0\left(\dfrac{P}{P_0}\right)^{2/(3\gamma-1)}.
\label{eqn:m2}
\end{align}
The period-derivative, $\dot{P}$, is
\begin{align}
\dfrac{\dot{P}}{P} &= \left(\dfrac{3\gamma-1}{2}\right)\dfrac{\dot{M}_{\rm d}}{M_{\rm d}} = - \left(\dfrac{3\gamma-1}{2}\right)\dfrac{\dot{M}}{M_{\rm d}},
\label{eqn:pdmd}
\end{align}
where we used the assumption of conservative mass transfer to obtain the last equation.


Next, we consider the evolution of the angular momentum of the binary. The spin-angular momenta of the donor and accretor are much lower than the orbital angular momentum of the system for all relevant values of the mass-ratio $q$ (even if the stars are tidally synchronized), and hence to a good approximation, the total angular momentum (and its evolution) is dominated by the orbital angular momentum:
\begin{equation}
J=\sqrt{\dfrac{Ga}{M}}M_{\rm a}M_{\rm d}=\sqrt{GaM_{\rm d}^3}\dfrac{\sqrt{\tilde{q}}}{q}.
\end{equation}
We use Eqn.~\eqref{eqn:roche} and the assumption that the donor fills its Roche-lobe to write the semi-major axis $a$ in terms of the donor radius $R_{\rm d}$. We then use the assumption of conservative mass transfer along with the mass-radius relation of Eqn.~\eqref{eqn:MR} and express the time evolution of the angular momentum as
\begin{equation}\label{eqn:jdot}
\dfrac{\dot{J}}{J}=\dfrac{1}{2}\dlndt{a}-\dfrac{3}{2}\dfrac{\dot{M}}{M_{\rm d}}+\dfrac{1}{2}\dlndt{\tilde{q}}-\dlndt{q}=\left(\dfrac{6q-3\gamma-5}{6}\right)\dfrac{\dot{M}}{M_{\rm d}}.
\end{equation}
Under our assumption that the angular momentum losses from the orbit are due to gravitational radiation\footnote{This may only be a valid approximation for AM CVns with accretion discs due to the more complicated angular momentum flux associated with direct accretion (see e.g. \citealt{V88}).}
\begin{align}
\nonumber
\dfrac{\dot{J}_{\rm GR}}{J}&=-\dfrac{32}{5}\dfrac{G^3}{c^5}\dfrac{MM_{\rm a}M_{\rm d}}{a^4}=-\dfrac{32}{5}\dfrac{G^3}{c^5}\dfrac{1}{q\tilde{q}}\dfrac{M_{\rm d}^3}{a^4}\\
&=-\dfrac{2\pi}{5}
\dfrac{\tilde{q}^{1/3}}{q}
\dfrac{1}{P_0}
\left(\dfrac{4\pi R_0}{\eta cP_0}\right)^{5}
\left(\dfrac{P}{P_0}\right)^{\frac{6-8 \gamma }{3 \gamma -1}}.
\label{eqn:jgr}
\end{align}
Combining Eqns.~\eqref{eqn:jdot} and \eqref{eqn:jgr}, we can directly relate the orbital period to the mass transfer rate of the binary
\begin{align}
\dot{M}
&=\dfrac{12\pi}{5}\dfrac{1}{(1-3\gamma)f\left(q,\gamma\right)}\dfrac{M_0}{P_0}\left(\dfrac{4\pi R_0}{\eta cP_0}\right)^{5}\left(\dfrac{P}{P_0}\right)^{8\left(\frac{1-\gamma}{3\gamma-1}\right)},
\label{eqn:mdot}
\end{align}
where the function $f$ is
\begin{equation}\label{eqn:fqg}
f\left(q,\gamma\right)\equiv\dfrac{q}{\tilde{q}^{1/3}}\dfrac{\left(5-6q+3\gamma\right)}{\left(1-3\gamma\right)}=\dfrac{q^{2/3}\left(1+q\right)^{1/3}\left(5-6q+3\gamma\right)}{\left(1-3\gamma\right)}.
\end{equation}
For a given/assumed $\gamma$, this result in combination with Eqns.~\ref{eqn:pdmd} and \ref{eqn:m2} can be used to infer the coefficient\footnote{$M_0$ and $R_0$ are not independent degrees of freedom; the quantity $R_0/M_0^\gamma$ is a single degree of freedom.} of the mass radius relation of the donor:
\begin{equation}
R_0=\dfrac{\eta}{4\pi}c P \left(\dfrac{c^3 M_0 P}{16\pi G}\right)^\gamma\left[\dfrac{5}{6\pi}\fqg\dot{P}\right]^{\frac{1-3\gamma}{5}}.
\end{equation}
Feeding this back into our previous calculations, we can reduce them to a final system of four equations:
\begin{align}\label{eqn:R_d}
\dot{M}&=\dfrac{1}{8\pi}\dfrac{c^3}{G}\dfrac{\left(\dot{P}\right)^{8/5}}{(1-3\gamma)}\left[\dfrac{5}{6\pi}\fqg\right]^{3/5},\\
\label{eqn:m_doner}
M_{\rm d}&=\dfrac{1}{16\pi}\dfrac{c^3}{G}P\left(\dot{P}\right)^{3/5}\left[\dfrac{5}{6\pi}\fqg\right]^{3/5}, \\
M_{\rm a}&=\dfrac{1}{q}M_{\rm d}, \, {\rm and} \\
R_{\rm d}&=\dfrac{\eta}{4\pi}c P \left[\dfrac{5}{6\pi}\fqg\dot{P}\right]^{1/5}.
\end{align}
Of the quantities on the right-hand side of the above equations, the period $P$, and the period derivative $\dot{P}$ are observable, while the exponent $\gamma$ can be determined by stellar modeling of the donor, leaving one free parameter: the mass-ratio $q$. 
Without further observational guidance, testing this model would involve comparing the predicted accretion rate $\dot{M}$, for a range of plausible values of $q$, to the value inferred from optical observations. 
If there is additional spectral information about the system, it becomes possible to use some velocities to further constrain the parameter space.
We now outline how this can be done.




The radius of the outer disk edge can be approximated as \citep{WAR03}
\begin{align}
    R_{\rm de}&=\dfrac{0.6}{1+q}a.
\end{align}
Assuming a Keplerian velocity profile, the orbital velocity of the outer disk edge is
\begin{align}
    v_{\rm de}&=\sqrt{\dfrac{GM_{\rm a}}{R_{\rm de}}}=0.495\dot{P}^{1/5}q^{1/6}\left[(1+q)\fqg\right]^{1/3},
\end{align}
where we used Eqns.~\ref{eqn:kep} and \ref{eqn:m_doner} to obtain the right hand side.
The semi-major axis of the accretor's orbit about the centre of mass is
\begin{align}
    a_{\rm a}&=a\dfrac{M_{\rm d}}{M_{\rm a}+M_{\rm d}}=\tilde{q}a,
\end{align}
making the orbital velocity of the accretor
\begin{align}
    v_{\rm a}&=a_{\rm a}\dfrac{2\pi}{P}=\dfrac{1}{2}\left(\dfrac{5}{6\pi}\right)^{1/5}c\dot{P}^{1/5}\left[\dfrac{q^2\fqg}{\left(1+q\right)^2}\right]^{1/3}.
\end{align}
Any observational inference of $v_{\rm de}$ or $v_{a}$ (e.g. via phase-resolved spectroscopy) will likely depend on inclination (as well as $\gamma$), however their ratio,
\begin{align}
\label{eqn:vratio}
    \dfrac{v_{\rm de}}{v_{\rm a}}&=1.29\dfrac{(1+q)}{\sqrt{q}}.
\end{align}
does not. Therefore it is possible to observationally determine the mass-ratio.
Once $q$ is determined, the inclination can also be inferred by comparing the observed line-of-sight projected velocities ($v_{\rm de,obs},\,v_{\rm a,obs}$) to the model velocities
\begin{align}
\label{eqn:inclination}
    i=\arcsin{\left[\dfrac{v_{\rm de}(q,\gamma)}{v_{\rm de,obs}}\right]}
    =\arcsin{\left[\dfrac{v_{\rm a}(q,\gamma)}{v_{\rm a,obs}}\right]}.
\end{align}

\section{Application to ES Ceti}

We shall now see how well the AM CVn system ES Ceti conforms to the model of Section~\ref{sec:model}. Here are the
relevant observables:
\begin{itemize}
	\item Orbital period: $P=620.211685(3)$ s \citep{ESP05,COP11}
	\item Period change: $\dot{P}=3.2(1)\times10^{-12}$ \citep{fdot_ES_Ceti}
	\item Bolometric flux: $F=9_{-4}^{+2}\times 10^{-10}$ erg s$^{-1}$ cm$^{-2}$ \citep{ESP05}\footnote{\citet{ESP05} state this flux as a luminosity scaled to a distance of 300 pc.}
	\item Parallax: $\varpi=0.60\pm0.1$ mas \citep{GaiaDR2}
	\item Phase-resolved spectroscopy \citep{spectrum}
\end{itemize}
We note that both the detection of the period change ($\dot{P}$) and the parallax ($\varpi$) are both recent observations. In fact before the release of Gaia DR2\footnote{\citet{KUP18} were first to note the updated Gaia parallax for ES Ceti.}, all inferences of $\dot{M}$ for ES Ceti assumed a distance of $d=350$ pc. With the measurement of the Gaia parallax it is clear that $d=1.7\pm0.3$~kpc.

\begin{figure}
	\includegraphics[width=.99\columnwidth]{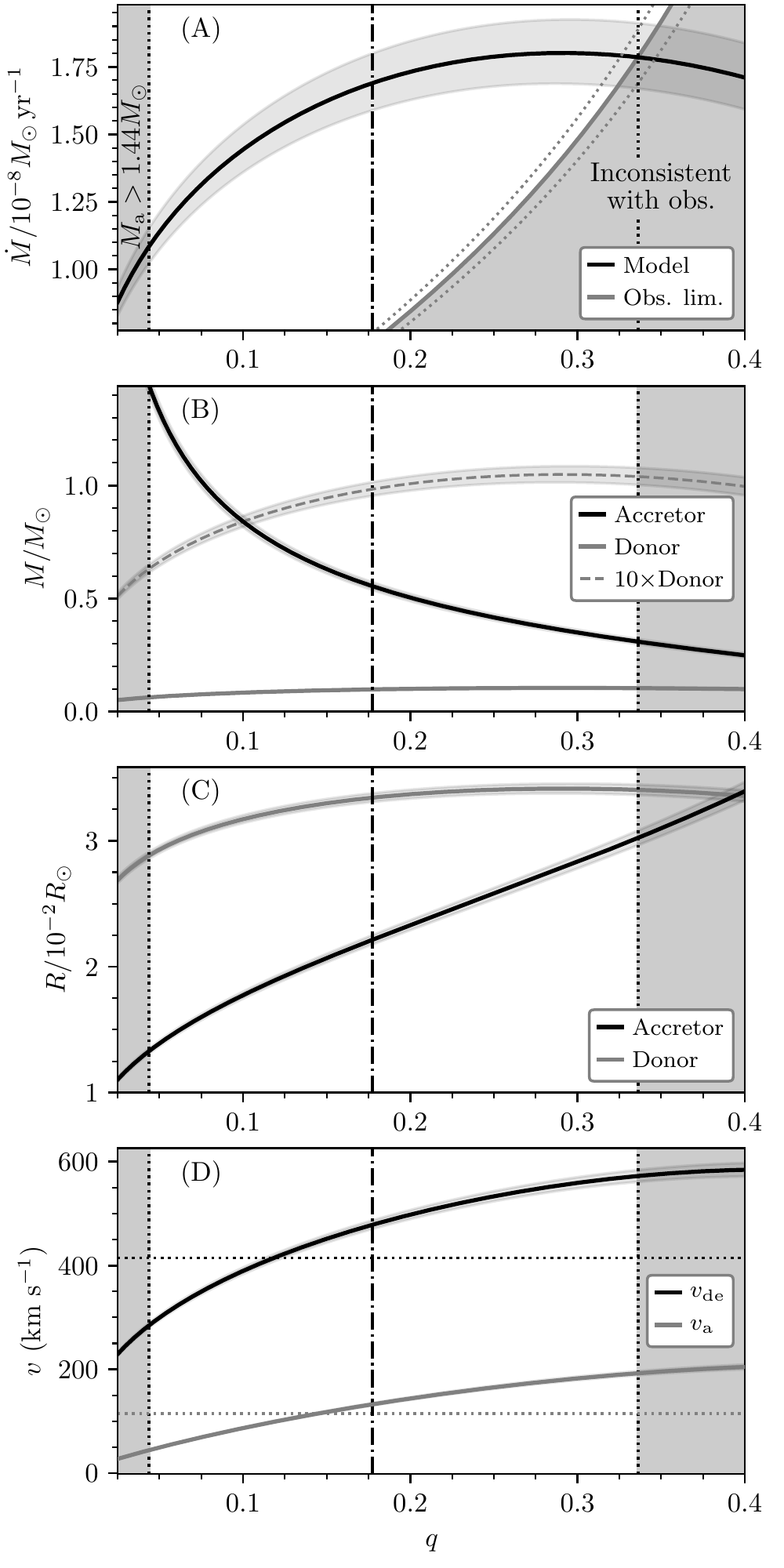}
	\vspace*{-.3cm}
	\caption{
		Various system parameters of ES Ceti (see Eqns.~\ref{eqn:mr1}-\ref{eqn:va_es_ceti}) as a function of mass-ratio $q$ for $\gamma=-0.30\pm0.02$ (see Eqn.~\ref{eqn:MR}). The vertical dashed-dotted line is our estimate of the mass-ratio. The lightly shaded region around each model line corresponds to the variation in $\gamma$. The $\gamma=-0.28$ contours extend to higher values for all curves with the exception of the predicted $\dot{M}$ in which the $\gamma=-0.32$ curve is above the fiducial line. 
		A) Mass accretion rate predicted by the model (black) and observational lower limit of accretion luminosity (grey); this lower limit corresponds to
		$F_{\rm obs}-2\sigma_{F}=1\times10^{-8}$ erg s$^{-1}$ cm$^{-2}$, 
		$\varpi{\rm obs}+2\sigma_\varpi=0.8$ mas, and the accretor properties plotted in the other panels. The right vertical black dotted line corresponds to the intersection of these curves, making the parameter space to the right of this line (shaded grey) inconsistent with the data.
		B) Masses of the accretor (black) and donor (solid grey), as well as donor mass scaled by 10 (dashed grey) to see its variation better. The left vertical black dotted line corresponds to $M_{\rm a}=1.44M_{\sun}$ (Chandrasekhar mass), making the parameter space to the left of this line (shaded grey) physically inconsistent. 
		C) Radii of the accretor assuming the mass radius relation Eqn.~\ref{eqn:mr1} (black) and donor (grey).
		D)  Orbital velocities of the disk edge and accretor. The horizontal dashed lines are the corresponding observationally estimated line-of-sight projected velocities.
	}
	\label{fig:param}
\end{figure}

Assuming that all the gravitational binding energy from all accreted material gets converted to photons then
\begin{align}
\nonumber
\dot{M}&=L\dfrac{R_{\rm a}}{GM_{\rm a}}=4\pi \dfrac{R_\star}{GM_\star}F \left(\dfrac{\varpi}{\text{mas}}\right)^{-2} \text{kpc}^2 \left(\dfrac{M_{\rm a}}{M_\star}\right)^{\delta-1}\\
\label{eqn:MdotObs}
&=6.29\!\times\! 10^{-8} M_{\sun}\text{ yr}^{-1}F_{-9}\left(\dfrac{\varpi}{0.6\text{ mas}}\right)^{-2}\left(\dfrac{M_{\rm a}}{0.8M_{\sun}}\right)^{-1.53},
\end{align}
where $F_{-9}$ is the flux scaled by $10^{-9}$ cgs units, and the following mass radius relation for the primary is used:
\begin{equation}\label{eqn:mr1}
R_{\rm a}=R_\star\left(\dfrac{M_{\rm a}}{M_\star}\right)^\delta=
1.8\times 10^{-2} R_{\sun} \left(\dfrac{M_{\rm a}}{0.8M_{\sun}}\right)^{-0.53}
\end{equation}
This assumed mass radius relation is a power law fit to the \citet{CAM17} results for a hydrogen-deficient white dwarf with $T_{c}=3\times10^7$ K, and $0.55<M_{\rm a}/M_{\sun}<0.95$. This central temperature is based on the \citet{BIL06} AM CVn accretor cooling track with an age roughly equal to $P/\dot{P}=6.14$ Myr.

By plugging in the observed values of $P$ and $\dot{P}$ into the equations in Section~\ref{sec:model} we get:
\begin{align}
\label{eqn:mdot_model}
\dot{M}&=4.7\pm0.2\times 10^{-8}M_{\sun}\text{ yr}^{-1}\dfrac{\left[\fqg\right]^{3/5}}{(1-3\gamma)}\\
M_{\rm d}&=0.144\pm0.003M_{\sun}\left[\fqg\right]^{3/5}\\
M_{\rm a}&=0.144\pm0.003M_{\sun}\dfrac{\left[\fqg\right]^{3/5}}{q}\\
\label{eqn:Rd}
R_{\rm d}&=3.79\pm0.02\times 10^{-2}R_{\sun}\left[\fqg\right]^{1/5}\\
v_{\rm de}&=746\pm75\text{ km s}^{-1}\left[(1+q)\fqg\right]^{1/3}q^{1/6}\\
\label{eqn:va_es_ceti}
v_{\rm a}&=578\pm19\text{ km s}^{-1}\left[\dfrac{q^2\fqg}{\left(1+q\right)^2}\right]^{1/3}
\end{align}
where $\fqg$ is defined in Eqn.~\ref{eqn:fqg} and the stated uncertainty primarily comes from the uncertainty in the measurement of $\dot{P}$.
We plot these equation for $\gamma=-0.3$ (consistent with a low mass white dwarf donor \citealt{DB03}) in Fig.~\ref{fig:param}. We note that the recovered $M_{\rm d}\approx0.08M_{\sun}$, $R_{\rm d}\approx0.03R_{\sun}$ are consistent with the white dwarf donor models derived by \citet{DB03} for a central temperature of $\sim 10^7$ K.

The functional form of Eqn.~\ref{eqn:mdot_model} and $\fqg$ admit a maximum possible accretion rate for each choice of $\gamma\neq1/3$, which can be determined by solving $\partial\dot{M}/\partial q=0$. The analytic expression for the general $\dot{M}_{\rm max}(\gamma)$ is both tedious to write and unilluminating so we only state the value for our fiducial $\gamma$:
\begin{align}
\dot{M}_{\rm max}\left(\gamma=-0.3\right)
=1.8\pm0.1\times 10^{-8} M_{\sun}\text{ yr}^{-1}.
\end{align}
While $1.1\lesssim\dot{M}/10^{-8}M_{\sun}\text{ yr}^{-1}\lesssim1.8$ is lower than the coefficient listed in Eqn.~\ref{eqn:MdotObs} it is still consistent with the observational data. To show this we plot the accretion rate for $F_{\rm obs}-2\sigma_{F}=1\times10^{-8}$ erg s$^{-1}$ cm$^{-2}$, 
$\varpi{\rm obs}+2\sigma_\varpi=0.8$ mas in Fig.~\ref{fig:param} (grey curve in top panel), where $\sigma_{F}$ and $\sigma_\varpi$ are the reported errors in the observed flux ($F_{\rm obs}$) and parallax ($\varpi_{\rm obs}$) respectively.
Over a range of values of the mass-ratio $q$, the values of $\dot{M}$ predicted by the gravitational-wave-moderated accretion model are consistent with those inferred from optical observations.

We can make further progress by using the \citet{spectrum} HeII 5411 \AA$\,$ observations. 
We use their images of phase resolved spectra to crudely estimate the velocities of the disk-edge and the accretor as
\begin{align}
    v_{\rm de,obs}&\approx415\text{ km s}^{-1}, \, {\rm and} \\
    v_{\rm a,obs}&\approx115\text{ km s}^{-1}.
\end{align}
The full data in \citet{spectrum} was not available and hence we cannot securely assign errors to these estimates.

Utilising Eqn.~\ref{eqn:vratio} we can estimate the mass ratio of ES Ceti:
\begin{align}
    q\approx0.18.
\end{align}
The dotted-dashed vertical line in Fig.~\ref{fig:param} marks this value of the mass-ratio. We note that all other parameters take very reasonable values for this value of the mass-ratio.

Additionally, using Eqn.~\ref{eqn:inclination}, we infer (assuming $\gamma=-0.3$) an inclination of
\begin{align}
    i\approx60^\circ.
\end{align}

\section{Discussion}

\begin{table*}
\footnotesize
\setlength{\tabcolsep}{4pt}
\begin{tabular}{lcccccc}
	\hline
	\hline \\[-2.3ex]
	& Outburst & {$P_{\rm orb}$} & {$\dot{M}_{\rm low}$} & {$\dot{M}$} & {$\dot{M}_{\rm high}$} & \\
	{System Name} & State & {(min)} & {$(10^{-9} M_{\sun} \,{\rm yr}^{-1})$} & {$(10^{-9} M_{\sun} \,{\rm yr}^{-1})$} & {$(10^{-9} M_{\sun} \,{\rm yr}^{-1})$} & {Source}\\
	\hline
	\hline \\[-2.3ex]
	\multicolumn{7}{c}{Persistent High State}\\
	\hline
ES Ceti       & H & 10.3 & 26.0 & 63.0 & 82.0 & \citet{ESP05}+Gaia \\
SDSSJ1351    & H & 15.7 & 0.2 & 2.6 & 35.5 & R+18 \\
AM CVn       & H & 17.1 & 5.6 & 7.1 & 9.3 & \citet{ROE07} \\
SDSSJ1908    & H & 18.1 & 354.8 & 660.7 & 1230.3 & R+18 \\
HP Lib       & H & 18.4 & 3.5 & 5.5 & 8.5 & R+18 \\
	\hline
	\multicolumn{7}{c}{Outbursting}\\
	\hline
ASASSN-14cc  & O & 22.5 & 1.7 & 2.3 & 3.2 & R+18 \\
CR Boo       & O & 24.5 & 0.4 & $0.7^*$ & 1.2 & \citet{ROE07} \\
KL Dra       & O & 25.0 & 0.5 & $0.9^*$ & 1.6 & \citet{RAM10} \\
V803 Cen     & O & 26.6 & 0.6 & $1.0^*$ & 1.6 & \citet{ROE07} \\
ASASSN-14mv  & O & 41.0 & 1.8 & 2.6 & 3.7 & R+18 \\
ASASSN-14ei  & O & 43.0 & 0.1 & 0.1 & 0.1 & R+18 \\
	\hline
	\multicolumn{7}{c}{Mix}\\
	\hline
SDSSJ1525    & L & 44.3 & 0.0 & 0.1 & 0.2 & R+18 \\
SDSSJ1411    & O & 46.0 & 0.0 & 0.0 & 0.1 & R+18 \\
GP Com       & L & 46.6 & 0.0 & 0.0 & 0.0 & R+18 \\
Gaia14aae    & O & 49.7 & 0.0 & 0.0 & 0.0 & R+18 \\
	\hline
	\multicolumn{7}{c}{Persistent Low State}\\
	\hline
SDSSJ1208    & L & 53.0 & 0.0 & 0.0 & 0.0 & R+18 \\
SDSSJ1137    & L & 59.6 & 0.0 & 0.0 & 0.0 & R+18 \\
V396 Hya     & L & 65.1 & 0.0 & 0.0 & 0.0 & R+18 \\
SDSSJ1319    & L & 65.6 & 0.0 & 0.0 & 0.0 & R+18 \\
\hline
\end{tabular}
\caption{Observed accretion rate and period of AM CVns. $P_{\rm orb}$ is the orbital period of the binary, $\dot{M}_{\rm low}$ and $\dot{M}_{\rm high}$ are the lower and upper bounds on the mass transfer rate respectively. $\dot{M}$ is the estimated mass transfer rate; values denoted by an asterisk are simply taken to be the geometric mean of the upper and lower bounds. The source R+18 is \citet{RAM18}.
}
\label{table:amcvn}
\end{table*}

\begin{figure}
	\includegraphics[width=1.0\linewidth]{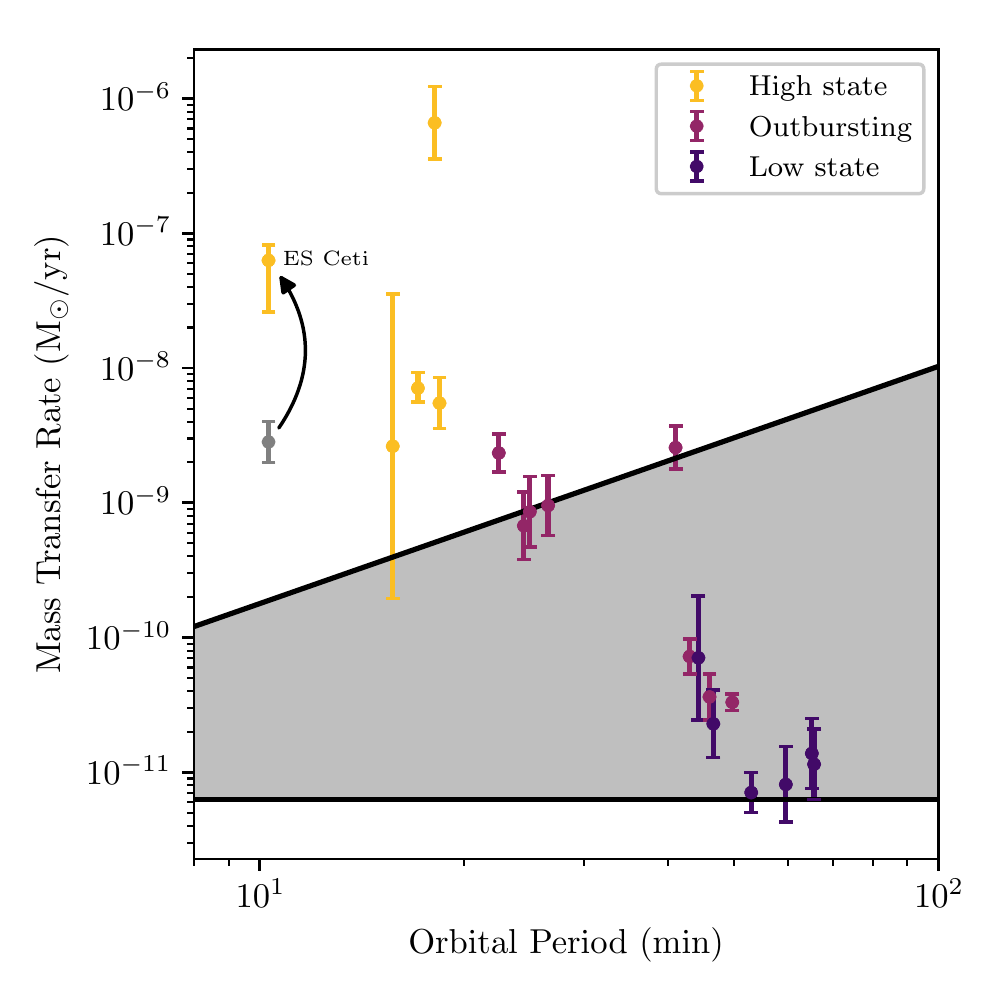}
	\caption{
		Limiting mass transfer rates for disc instability within AM CVn accretion discs as a function of orbital period. Solid black lines are limiting transfer rates from \citet{COL18}, assuming a local instability must occur between the outer disc edge and $4R_{\rm a}$, for $M_{\rm a}=1M_{\sun}$ (see Eqn.~\ref{eqn:mr1}).  This choice of $4R_{\rm a}$ is ad hoc and chosen to based on the observations.
		The data for several observed AM CVn with observationally inferred mass transfer rates is also plotted (see Table~\ref{table:amcvn}). The colouring of these points corresponds to the observed state of the AM CVns with persistent high state systems in yellow, outbursting systems in magenta, and persistent low state systems in dark-purple. We also show ES Ceti's inferred accretion rate prior to the Gaia parallax measurement (grey point).
	}
	\label{fig:obs}
\end{figure}

AM~CVns exhibit similar accretion disc variability
phenomenology of hydrogen-dominated 
accretion discs around white dwarfs (i.e. cataclysmic variables), including
persistent and outbursting systems, superhumps,
quasiperiodic oscillations (QPOs),
and broadband noise (e.g. \citealt{RAM12,CAM15,KUP15,LEV15,COL18}).
The outbursting systems are generally
dominated by superoutbursts, although normal outbursts have also been observed
\citep{LEV11}.
In the past, changes in measured distances to accreting white dwarfs have changed the predicted activity of the accretion disc; in the historical case of the dwarf nova SS Cygni, the change in measured parallax caused the inferred accretion rate to cross the theoretical critical value for disc instabilities. With the current accepted measurement placing SS Cygni in the outbursting parameter space consistent with its observed accretion disc outbursts
\citep[see e.g.][]{2007A&A...473..897S,2013Sci...340..950M,2013ApJ...773L..26N}. While the inferred accretion rate for ES Ceti is increased by an order of magnitude, this change does not cross the critical values determined by \citet{COL18} (see Fig.~\ref{fig:obs}), maintaining consistency with the lack of observed outbursts in ES Ceti. This new accretion rate also makes the relation between $P$ and $\dot{M}$ in observed AM~CVns more consistent with that of a power-law (see Table~\ref{table:amcvn} and Fig.~\ref{fig:obs}). 

The new inferred accretion rate is on the high side, but still consistent with our model. As the Gaia parallax becomes more precise it may become necessary to consider additional sources of orbital angular momentum losses (e.g. spin-up of the accretor) or mass-loss (e.g. accretion disc winds). We also suggest that the spectral data on ES Ceti be reanalysed with the knowledge of the Gaia parallax, as this may result in a more precise and accurate determination of the accretion rate.

One could ask the question why can this exercise only be done with ES Ceti? In short because ES Ceti is the shortest period AM CVn with an accretion disc. The expectation is that $P/\dot{P}$ is shorter for tighter systems, making $\dot{P}$ easier to measure. Additionally, having an accretion disc means that a higher fraction of the accretion luminosity is emitted in optical light, making the observations accessible to amateur astronomers whose data were instrumental in measuring the $\dot{P}$ of ES Ceti. Accordingly, ES Ceti is the first AM CVn that is expected to have the requisite measurements.

ES Ceti is also a verification binary for LISA with a potential signal-to-noise ratio (SNR) of $154\pm 2$ after 4 years of observations \citep{KUP18}. 
We now estimate whether LISA can, over this timescale, detect the period derivative due to the orbital evolution of ES Ceti.

In the high SNR limit, the minimum detectable frequency-derivative ($\dot{f}$) would enhance the matched-filtering ${\rm SNR}^2$ by unity\footnote{A technical point is that we need to orthogonalise the effects of $\dot{f}$ with those of frequency $f$ and phase; we achieve this by measuring the frequency at the midpoint of the interval, and looking at the real part of the overlap.}. Alternatively, matched-filtering with templates without $\dot{f}$ would cause us to lose one unit of ${\rm SNR}^2$. Writing down the matched-filtering integral in the  time domain, we get
\begin{align}
\nonumber
    1-\dfrac{1}{{\rm SNR}^2} &= \dfrac{{\rm Re} \int_{-t/2}^{t/2}\id \tau \exp\left(i\delta \dot{f}\tau^2\right)}{\int_{-t/2}^{t/2}\id \tau}\\
    \nonumber
    &\approx \dfrac{{\rm Re} \int_{-t/2}^{t/2}\id \tau \left[1+i\delta\dot{f}\tau^2-\frac{1}{2}\left(\delta\dot{f}\tau^2\right)^2 \right]}{t}\\
    &= 1-\dfrac{\delta\dot{f}^2 t^4}{160},
\end{align}
where $\delta\dot{f}$ is the minimum detectable gravitational wave frequency derivative, and $t$ is the cumulative observing time. The solution for $\delta\dot{f}$ is
\begin{align}
    \delta\dot{f} &= \dfrac{4\sqrt{10}}{{\rm SNR}\times t^2} = 1.63\times10^{-10}\text{ Hz yr}^{-1} \left(\dfrac{\rm SNR}{154}\right)^{-1}\left(\dfrac{t}{4\text{ yr}}\right)^{-2}.
\end{align}
This can also be stated in terms of a minimum detectable orbital period change ($\delta\dot{P}$), as follows:
\begin{align}
    \delta\dot{P}&=\dfrac{\delta\dot{f}P^2}{2} = 9.93\times 10^{-13}\left(\dfrac{\rm SNR}{154}\right)^{-1}\left(\dfrac{t}{4\text{ yr}}\right)^{-2}\left(\dfrac{P}{620\text{ s}}\right)^2,
\end{align}
where the factor of two is due to the relationship between the gravitational-wave and the orbital frequencies. Comparison with the measured value of $\dot{P} = 3.2 \times 10^{-12}$ \citep{fdot_ES_Ceti} indicates that the widening of ES Ceti will be detectable by LISA.

We can also predict the gravitational wave amplitude of ES Ceti:
\begin{align}
    A&=\dfrac{2(G\mathcal{M})^{5/3}}{c^4 d}(\pi f)^{2/3}
     =\dfrac{5cP\dot{P}}{192d\pi^2}\dfrac{5-6q+3\gamma}{1-3\gamma}\\
     &=3.06\pm{0.1}\times 10^{-23}\left(\dfrac{\varpi}{0.60\text{ mas}}\right)\left(\dfrac{5-6q+3\gamma}{1-3\gamma}\right)\\
     \label{eqn:strain}
     &=6.61\pm{1.1}\times 10^{-23}\left(1-1.46q\right)\;\;(\text{for }\gamma=-0.3).
\end{align}
In the first equation, $\mathcal{M}$ is the chirp-mass of the binary, defined as $\mathcal{M} = M_{\rm d} / (q^2 (q + 1))^{1/5}$. In Eqn.~\eqref{eqn:strain} above, we have assumed a value of $\gamma=-0.3$, utilised the observed parallax and have propagated the errors on $\varpi$ (which dominate the uncertainty). This value is slightly lower than the strain \citet{KUP18} predict for ES Ceti, but consistent to within $2\sigma$ for all reasonable values of $q$. Hence, Eqn.~\ref{eqn:strain} suggests that the amplitude measured by LISA can be used to constrain the mass ratio of ES Ceti.

Note that the gravitational waveform does not evolve in the standard manner expected for point masses evolving purely due to GW emission ($\dot{f} \sim \mathcal{M}^{5/3} f^{11/3}$); in fact the signal slowly reduces in frequency over time.


\section{Conclusions}

The new Gaia parallax brings ES Ceti into better agreement with the model of gravitational-wave-moderated accretion, due to the increase in inferred accretion rate. With this measurement taken into account we find the observed orbital period, period derivative, and mass accretion rate are constant with our model. This model assumes the mass transfer is conservative (i.e. no mass is lost from the system) and that the only change of orbital angular momentum comes from the emission of gravitational radiation. Additionally, we assumed the mass radius relation of the accreting white dwarf is given by Eqn.~\ref{eqn:mr1}. This work finally confirms the theory of gravitational-wave-regulated accretion proposed by \citet{P67}.

\section*{Acknowledgements}

We thank Tomas Kupfer, Evan Bauer, Omer Blaes and Roman Rafikov for their useful discussions and insight generated from their work.
MC gratefully acknowledges support from the Institute for Advanced
Study, NSF via grant AST-1515763, and NASA via grant
14-ATP14-0059.
TV acknowledges support by the Friends of the Institute for Advanced Study. 
BZ acknowledges the support of The Peter Svennilson Membership
fund.
This work has made use of a single datum from the European Space Agency (ESA) mission
{\it Gaia} (\url{https://www.cosmos.esa.int/gaia}), processed by the {\it Gaia}
Data Processing and Analysis Consortium (DPAC, \url{https://www.cosmos.esa.int/web/gaia/dpac/consortium}). Funding for the DPAC
has been provided by national institutions, in particular the institutions
participating in the {\it Gaia} Multilateral Agreement.




\bibliographystyle{mnras}
\bibliography{citations}



%
%


\bsp	
\label{lastpage}
\end{document}